\documentclass[usenatbib,usehyperref]{mn2e}

\IfFileExists{srcltx.sty}{\usepackage[active]{srcltx}}

\usepackage{graphicx,amssymb,epstopdf,times,url} %
\usepackage{hyperref,breakurl}

 \def\fermi{\emph{Fermi}}

\usepackage[a4paper,totalwidth=480pt, totalheight=680pt]{geometry}

\title[Emission from the GC with Fermi]{A comment on the emission from the
  Galactic center as seen by the Fermi telescope} \author[A.~Boyarsky,
D.~Malyshev, O.~Ruchayskiy]{Alexey Boyarsky$^{1,2}$, Denys Malyshev$^3$, Oleg
  Ruchayskiy$^4$\\
  $^1$Ecole Polytechnique F\'ed\'erale de Lausanne, FSB/ITP/LPPC, BSP
  CH-1015, Lausanne, Switzerland\\
  $^2$Bogolyubov Institute for Theoretical Physics, Metrologichna str.,
  14-b, Kiev 03680, Ukraine\\
  $^3$Dublin Institute for Advanced Studies, Astronomy \& Astrophysics Section, 31 Fitzwilliam Place, Dublin, 2 Ireland\\
  $^4$CERN TH-Division, PH-TH, Case C01600, CERN, CH-1211 Geneva 23,
  Switzerland} \date{CERN-PH-TH-2010-291} 
\begin{document}

% \keywords{}
\maketitle
\begin{abstract}
  In the recent paper of \citet{Hooper10} it was reported that $\gamma$-ray
  emission from the Galactic Center region contains an excess compared to the
  contributions from the large-scale diffuse emission and known point sources.
  This excess was argued to be consistent with a signal from annihilation of
  Dark Matter with a power law density profile.  We reanalyze the \fermi\ data
  and find instead that it is consistent with the ``standard model'' of
  diffuse emission and of known point sources.  The main reason for the
  discrepancy with the interpretation of \citet{Hooper10} is different (as
  compared to the previous works) spectrum of the point source at the Galactic
  Center assumed by \citet{Hooper10}. We discuss possible reasons for such an
  interpretation.
\end{abstract}

\section{Introduction}
The origin of the emission from the Galactic Center (GC) at keV--TeV energies
has been extensively discussed in the literature over last few years.  In
their recent paper, \citet{Hooper10} claimed that the $\gamma$-ray emission
from the Galactic Center region, measured with the \fermi\ LAT
instrument~\citep{FermiOverview} cannot be described by a combination of
spectra of known point sources, diffuse emission from the Galactic plane and
diffuse spherically symmetric component (changing on the scales much larger
than $1^\circ$). An additional spherically symmetric component was suggested
to be needed in the central several degrees. This component was then
interpreted as a dark matter annihilation signal with the dark matter
distribution having power law density profile $\rho(r) \propto r^{-\alpha}$,
$\alpha \approx 1.34$.  The observed excess is at energies between $\sim
600$~MeV and $\sim 6$~GeV and the mass of the proposed DM particle was
suggested to be in the GeV energy band.

In this work we analyze the \fermi\ data, used in \citet{Hooper10}, utilizing
the data analysis tool, provided by the \fermi\ team.

\section{Data}

For our analysis we consider 2 years of \fermi\ data collected between August,
4th, 2008 and August 18th, 2010. The standard event selection for source
analysis, resulting in the strongest background-rejection power (\emph{diffuse
  event class}) was applied.\footnote{See e.g.
  \url{http://fermi.gsfc.nasa.gov/ssc/data/analysis/scitools}} In addition,
photons coming from zenith angles larger than $105^\circ$ were rejected to
reduce the background from gamma rays produced in the atmosphere of the Earth.

The \fermi's point-spread function (PSF) is non-Gaussian and strongly depends
on energy \citep{FermiCalibration,FermiOverview}. In order to properly take it
into account and better constrain the contributions from Galactic and
Extragalactic diffuse backgrounds we analyze a $10^\circ \times 10^\circ$
region around the Galactic Center.

\subsection{Model}
\label{sec:Model}

To describe emission in the $10^\circ \times 10^\circ$ region we use the model
containing two components -- point sources and diffuse backgrounds.

To model the contribution from the point sources we include 19 sources from 11
months \fermi\ catalog \citep{1FGLcat} falling into the selected region plus 4
additional sources described in \citet{Chernyakova10}. We fix the positions of
the sources to coordinates given in the catalog. We model their spectra as
power law (in agreement with \citealt{1FGLcat}). Thus we have 46 free
parameters (power law index and norm for each of the sources) to describe the
point-source component of the model.

To describe the diffuse component of emission, we use the models for the
Galactic diffuse emission (\texttt{gll\_iem\_v02.fit}) and isotropic
(\texttt{isotropic\_iem\_v002.txt}) backgrounds that were developed by the LAT
team and recommended for the high-level
analysis~\citep{FermiEGB}\footnote{\url{http://fermi.gsfc.nasa.gov/ssc/data/analysis/scitools/likelihood\_tutorial.html}}.
These models describe contributions from galactic and extragalactic diffuse
backgrounds correspondingly. The number of free parameters for the diffuse
background model is 2 (the norms for each of the backgrounds).  The total
number of free parameters in our model is thus 48.

This model is similar to the one described in \citet{Chernyakova10}.

\subsection{Analysis}
\label{sec:analysis}

The data analysis was performed using the LAT Science Tools package with the
P6\_V3 post-launch instrument response function \citep{Rando09}.

We find the best-fit values of all parameters of the model of
Section~\ref{sec:Model} (using \texttt{gtlike} likelihood fitting tool) and
determine resulting log-likelihood \citep{Mattox96} of the model.  Best fit
values for the obtained fluxes agree within statistical uncertainties with
fluxes reported in \fermi\ Catalog~\citep{1FGLcat} and in
\citet{Chernyakova10} (e.g. for the central source we obtained the flux
$5.68\times 10^{-8}~\mathrm{cts/cm^2/s}$ while the catalog gives
$(5.77\pm0.3)\times 10^{-8}~\mathrm{cts/cm^2/s}$).

\begin{figure}
  \includegraphics[width=.5\textwidth]{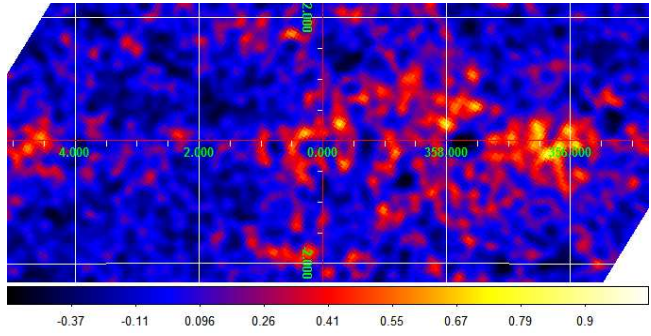}
  \caption{The map of significance of residuals for the region around the
    Galactic Center.}
\label{fig:rel_residuals}
\end{figure}

We then freeze the values of the free parameters of our model and simulate
spatial distribution of photons at energies above 1~GeV (using
\texttt{gtmodel} tool).  The significance of residuals, (Observation - Model)/
statistical error, is shown in Fig~\ref{fig:rel_residuals}. We see the absence
of structures in the central $2^\circ$ region.  The average value of residuals
is about 10\% in the $2^\circ$ region around the GC, compatible with estimated
systematic errors (10-20\%) of \fermi\ LAT at 1~GeV.\footnote{See e.g.
  \url{http://fermi.gsfc.nasa.gov/ssc/data/analysis/LAT\_caveats.html}}

Thus we see that the adopted model (point sources plus galactic and
extragalactic diffuse components) explains the emission from the GC region and
no additional components is required.

\begin{figure}
  \includegraphics[width=.5\textwidth]{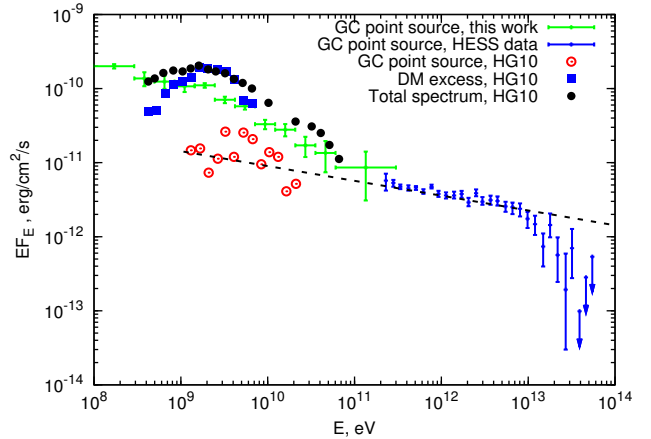}
  \caption{Spectrum of the point source at the GC reported in
    \citet{Chernyakova10} (green points) together with the HG10 total spectrum
    from $1.25^\circ$ (black points), excess (blue squares) and GC point
    source flux from HG10 (red open circles).  Continuation of the HESS data
    ~\citep{vanEldik:07,Aharonian:04} (blue points) data with a power law is
    shown with dashed black line.}
\label{fig:gc_spectrum}
\end{figure}

\begin{figure*}
  \centering
\begin{minipage}{.5\textwidth}
  \includegraphics[width=\linewidth]{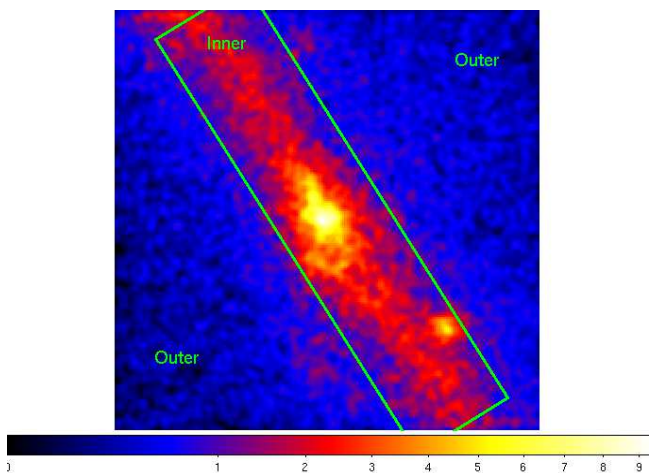}
\end{minipage}~\begin{minipage}{.5\textwidth}
  \includegraphics[width=\linewidth]{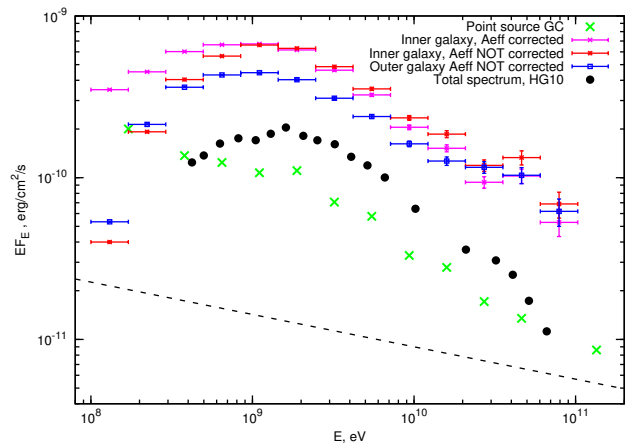}
\end{minipage}
\caption{\textbf{Left:} the ``inner'' ($5^\circ$ around the Galactic plane)
  and ``outer'' regions.  \textbf{Right:} Effects of the energy dependence of
  the effective area for the spectra of the ``inner'' and ``outer'' regions.
}
  \label{fig:aeff}
\end{figure*}

\section{Discussion}

We conclude that the signal within central $1^\circ{-}2^\circ$, containing the
``excess'' found by \citealt{Hooper10} (\textbf{HG10} hereafter), can be well
described by our model : (point sources plus Galactic and extragalactic
diffuse background components). The discrepancy is then due to a different
interpretation of the data.

The spectrum of the central point source (1FGL J1745.6-2900c, probably
associated with the Galactic black hole Sgr A$^*$) was taken in HG10 to be a
featureless power-law starting from energies about 10~TeV (results of HESS
measurements, blue points with error bars in Fig.~\ref{fig:gc_spectrum},
\citep{Aharonian:04,vanEldik:07}) and continuing all the way down to $\sim
1$~GeV. The flux attributed in this way to the central point source is
significantly weaker than in the previous works. For comparison, the (PSF
corrected) spectrum of the GC point source reported in~\citet{Chernyakova10}
is shown in Fig.~\ref{fig:gc_spectrum} in green points.  Its spectral
characteristics are fully consistent with the results of 11-months \fermi\
catalog~\cite{1FGLcat} ($\sim 6\times 10^{-8}~\mathrm{cts/cm^2/s}$ above
1~GeV, compared to the $\sim 5\times 10^{-9}~\mathrm{cts/cm^2/s}$ at the same
energies in HG10).  The change of the slope of the source spectrum below $\sim
100$~GeV, as compared with the HESS data is explained by \citet{Chernyakova10}
with the model of energy dependent diffusion of protons in the few central
parsecs around the GC.  Alternatively, the spectrum can be explained with the
model developed in \citet{Aharonian05}. The low-energy (GeV) component of the
spectra in this model is explained by synchrotron emission from accelerated
electrons, while high-energy (TeV) one by inverse Compton radiation of the
same particles.  According to the analysis of \citet{1FGLcat,Chernyakova10}
the central point source provides significant contribution to the flux in the
1.25$^\circ$ central region.  HG10 suggest, apparently, a different
interpretation.  They assume that there is no significant change in the
spectrum of the central source at $\sim 100$~GeV and the spectrum observed by
HESS at high energies continues to lower energies.  Then, large fraction of
the flux between the energies $\sim 600$ MeV and $\sim 6$~GeV has to be
attributed to the ``DM excess''. One of the reasons in favor of such an
interpretation could be the feature in the total spectrum from the central
region (rise between $\sim 600$~MeV and several GeV) discussed in HG10.  Such
a feature would also be consistent with a possible contribution from
millisecond pulsars \citep{Abazajian10a}, that is also expected to have a
maximum at $\sim 2-3$~GeV.

To illustrate the nature of the spectral shape at these energies we collected
``front converted'' (\textsc{front}) photons from the region of the width
$5^\circ$ around the Galactic Plane (the ``\textit{inner}'' region) and from
the ``\textit{outer}'' region as demonstrated on the left panel in
Fig.~\ref{fig:aeff}. The count rate from each of these regions was divided by
the \emph{constant} effective area ($3500~\mathrm{cm}^2$) to obtain the
flux.\footnote{The effective area of \fermi\ LAT is strongly energy dependent.
  The number $3500~\mathrm{cm}^2$, roughly corresponding to the effective area
  at $\sim 1$~GeV, is used here as a quick expedient (see below).}  One sees
that the total emission from both regions demonstrates the same spectral
behavior as the excess of HG10, suggesting that this spectral shape is
\emph{not} related to the physics of the several central degrees.  This drop
of flux at low energies is mainly due to the decreasing effective area of the
satellite.\footnote{\url{http://www-glast.slac.stanford.edu/software/IS/glast_lat_performance.htm}}
If we properly take into account the dependence of the effective area on
energy, we obtain the spectrum that ``flattens'' at small energies and exceeds
by a significant factor the flux from the central point source (as it should)
(compare red and magenta points on the right panel in Fig.~\ref{fig:aeff}).

Another reason for the decrease of the HG10 spectrum is the increase of
\fermi\ LAT PSF at low ($\lesssim 1$ GeV) energies.\footnote{For example, for
  normal incidence 95\% of the photons at $1$ GeV are contained within $\sim
  1.6^\circ$ and in $2.8^\circ$ at $500$ MeV} This means that if one collects
photons from a relatively small region, such that a contribution from its
boundary (with the PSF width) is comparable to the flux from the whole region,
the spectrum would artificially decline, due to increasing loss of photons at
low energies.  To disentangle properly what photons in the PSF region had
originated from a localized source, and what are parts of the diffuse
background, special modeling is needed. In the monotonic spectrum of the GC,
obtained by~\citet{Chernyakova10} both these effects (effective area and PSF)
were taken into account as it was obtained from $10^\circ\times10^\circ$
region, using the \fermi\ software.

\begin{figure*}
  \centering
  \begin{minipage}{.5\textwidth}
    \includegraphics[width=\linewidth]{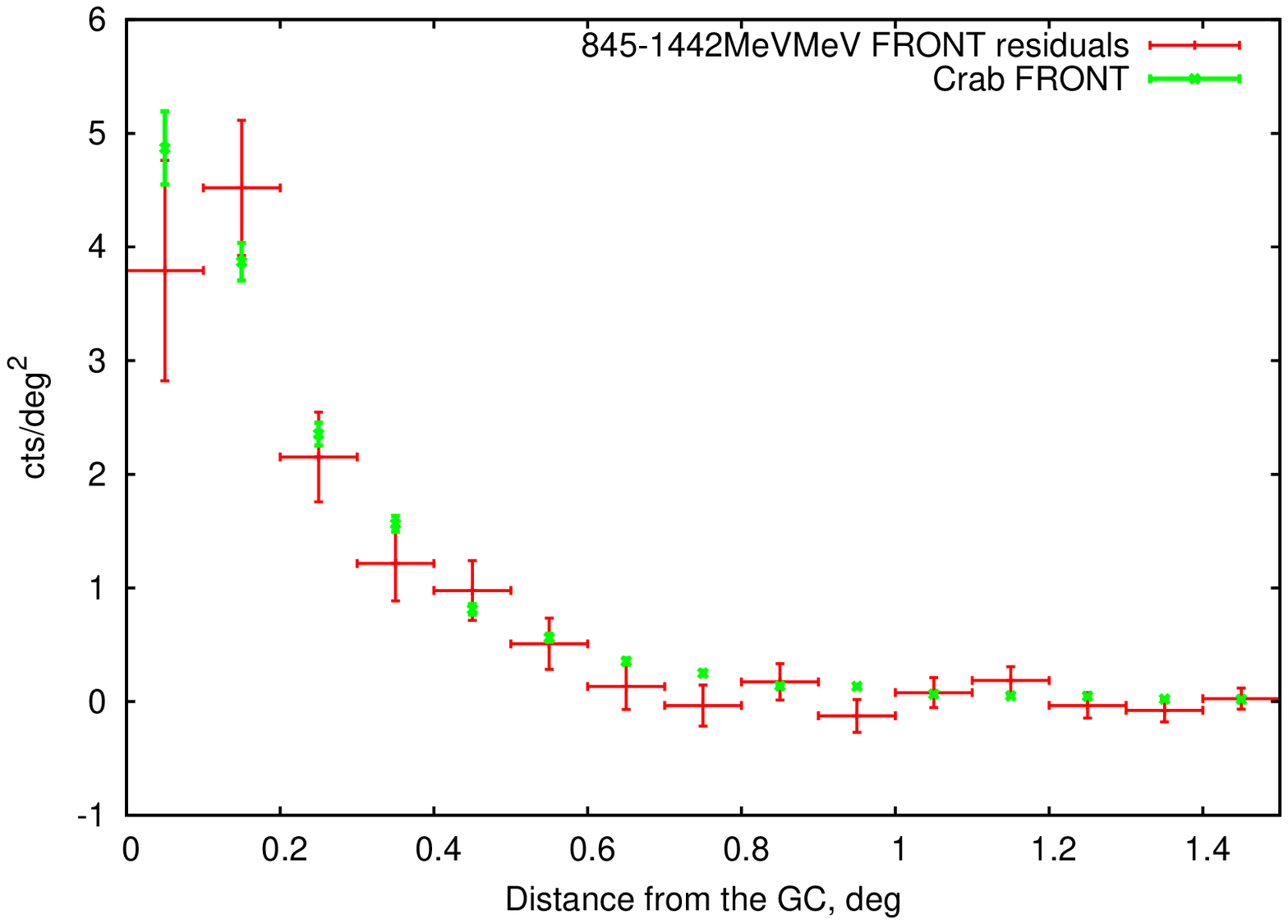}
  \end{minipage}~\begin{minipage}{.5\textwidth}
    \includegraphics[width=\linewidth]{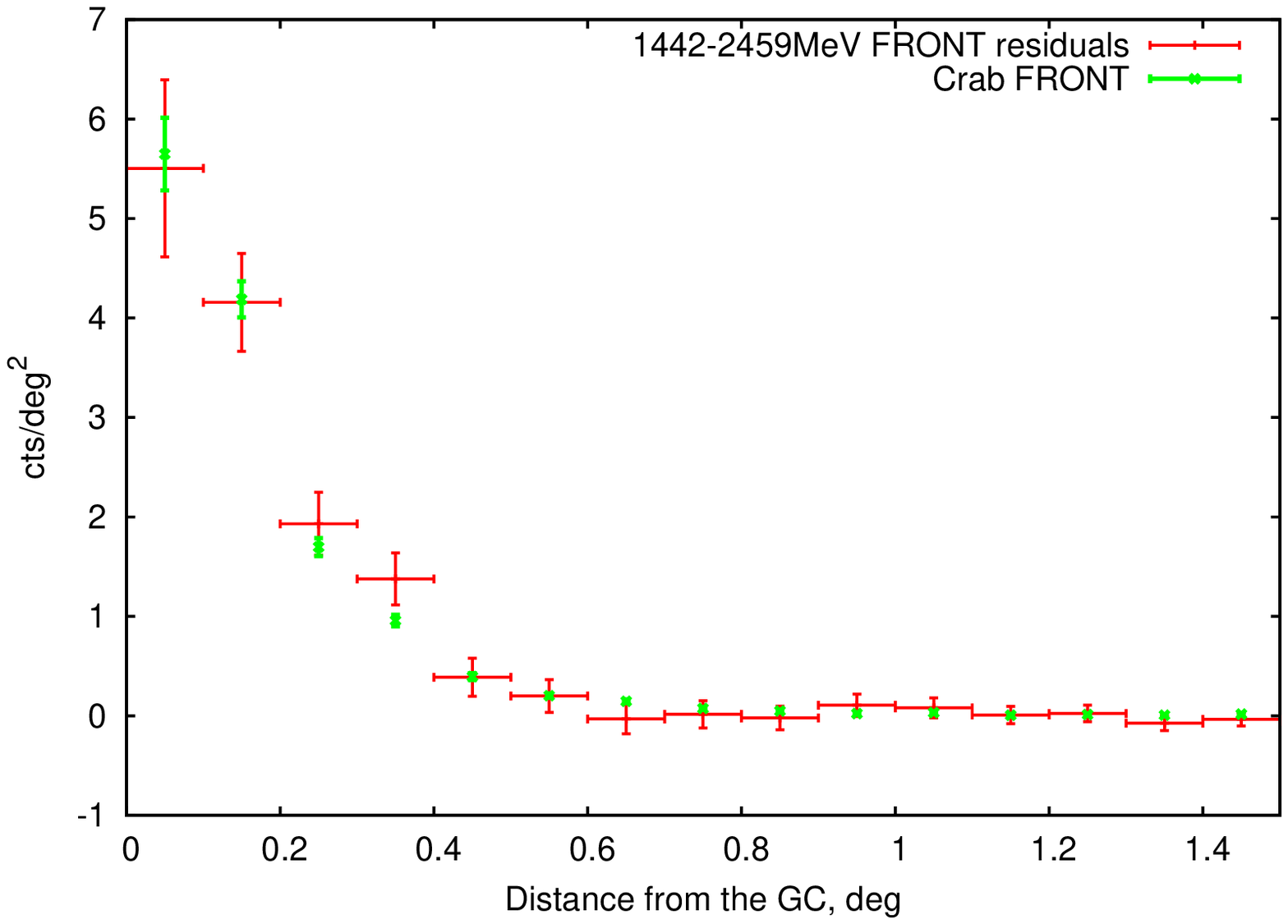}
  \end{minipage}
  \begin{minipage}{.5\textwidth}
    \includegraphics[width=\linewidth]{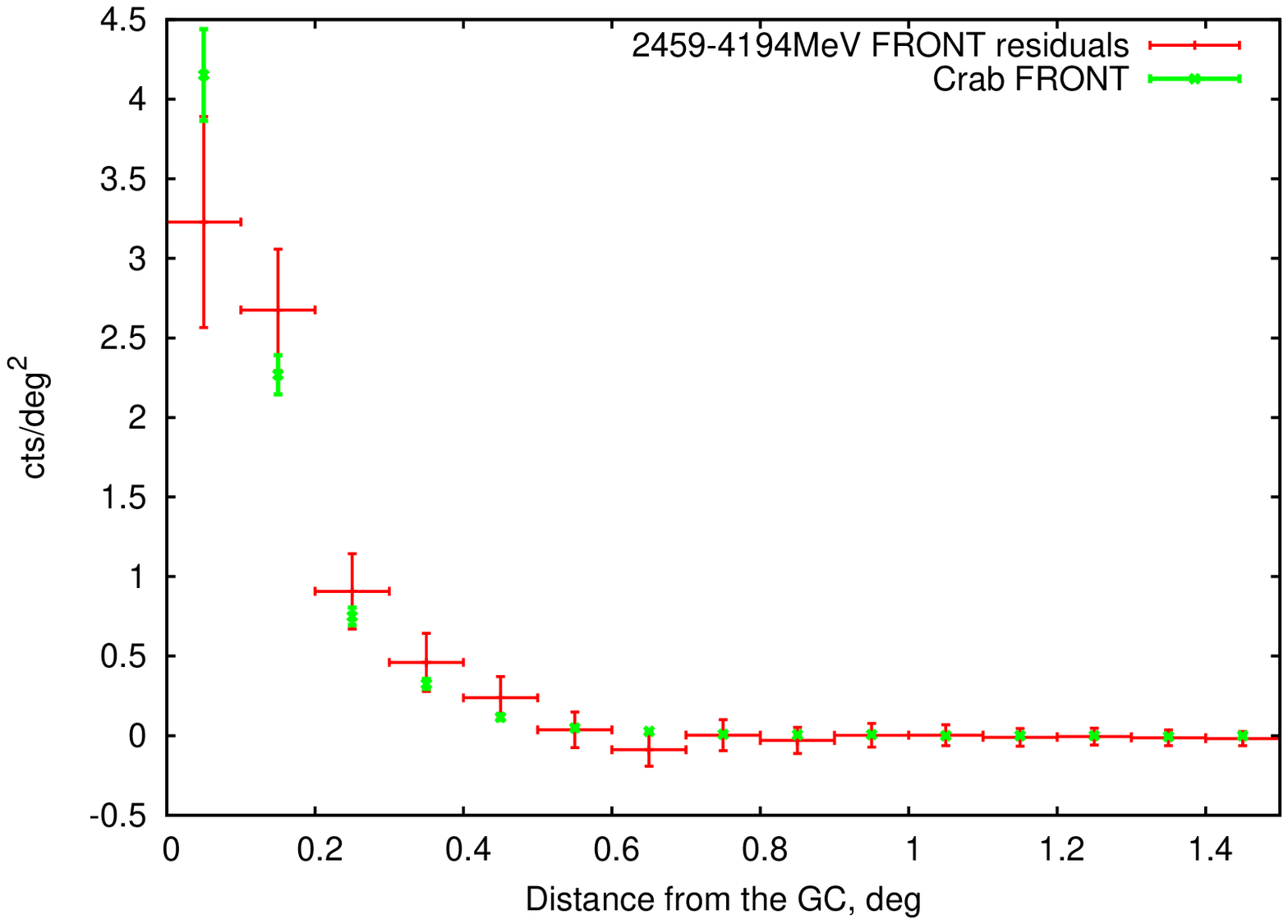}
  \end{minipage}~\begin{minipage}{.5\textwidth}
    \includegraphics[width=\linewidth]{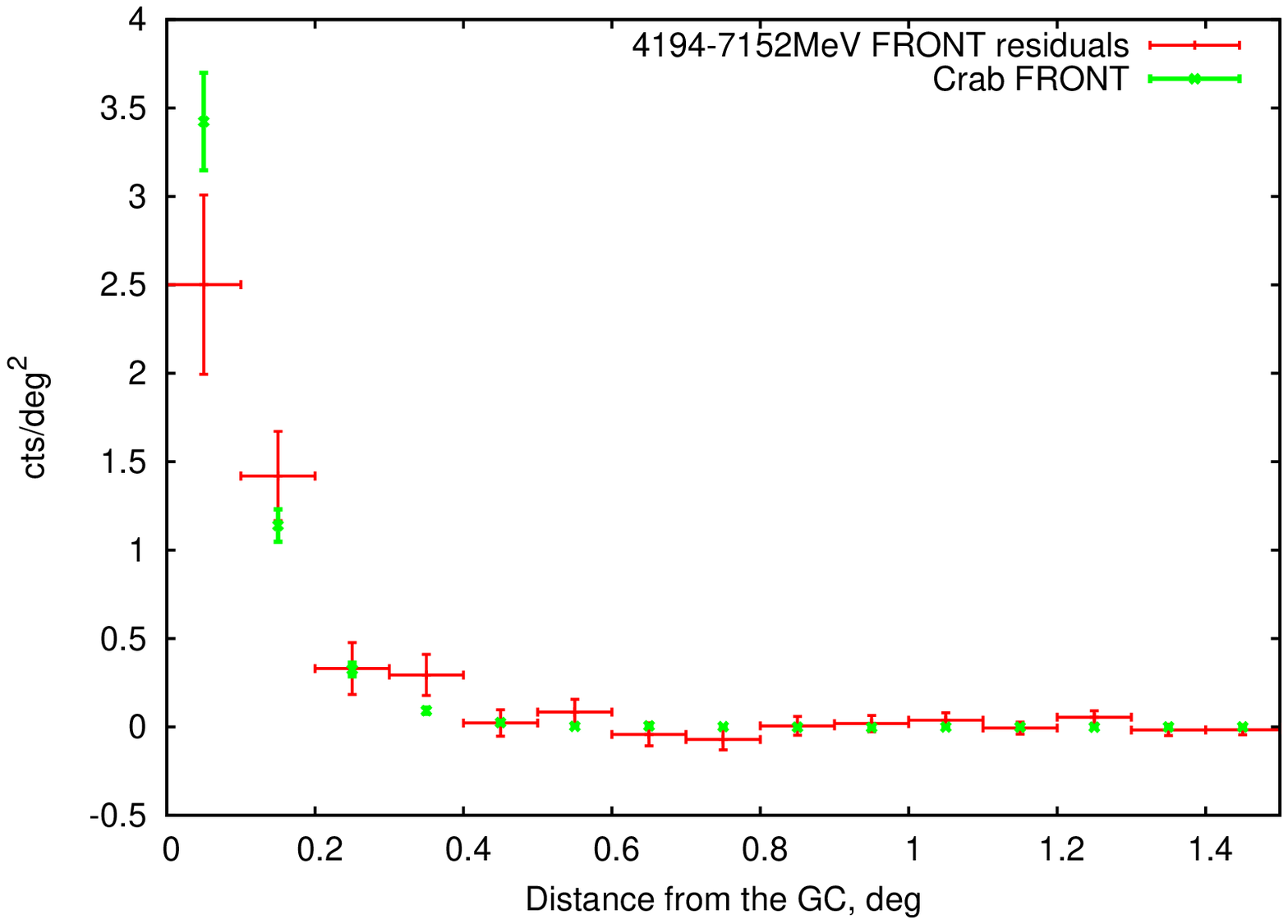}
\end{minipage}
\caption{Radial profile of residuals at different energies around the GC as
  compared to the radial profile of Crab emission (renormalized so that the
  total flux in each energy range coincide). In both cases only \textsc{front}
  photons were used.}
  \label{fig:crab}
\end{figure*}

To further check the nature of the emission from the central several degrees,
we took a fiducial model, that contained the same galactic and extragalactic
diffuse components plus all the same point sources, but \emph{excluding the
  point source in the center}. We then fit our data to this new model. Such a
fit attempts to attribute as many photons as possible from the region around
the GC to the emission of diffuse components.  The procedure leaves strong
positive residuals within the central $1{-}2^\circ$.  The spectrum of these
residuals is consistent with the spectrum of the central point source
of~\citet{Chernyakova10} (green points in Fig.~\ref{fig:gc_spectrum}). To
demonstrate, that the spatial distribution of these residuals is fully
consistent with the PSF of \fermi, we compare their radial distribution in
various energy bins with the radial distribution around the Crab pulsar (as it
was done e.g. in~\citet{Neronov:10b}). The pulsar wind nebula, associated with
the Crab has an angular size $\sim 0.05^\circ$~\citep{Hester:08}. Thus, for
\fermi\ LAT Crab is a point source. The radial profile of residuals at all
energies has the same shape as Crab, as Fig.~\ref{fig:crab} clearly
demonstrates. As an additional check, we repeated the above test using only
\textsc{front} photons (as in this case the PSF is more narrow) and arrived to
the same conclusion.

The above analysis demonstrates that the emission around the GC in excess of
diffuse components (galactic and extragalactic) is fully consistent with being
produced by the point source with the power-law spectrum, obtained
in~\cite{1FGLcat,Chernyakova10}, \emph{and no additional component is
  required.}

\begin{figure}
  \centering
  \includegraphics[width=\linewidth]{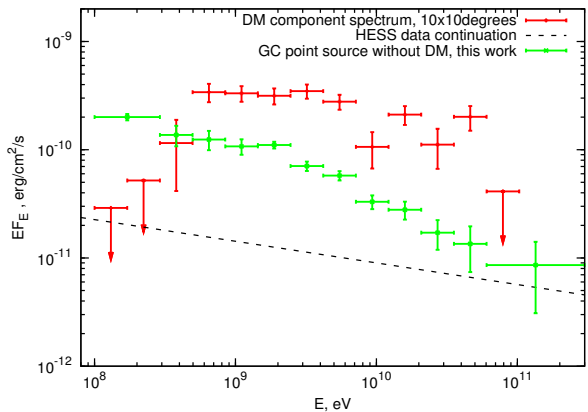}
  \caption{Spectrum of an additional spherically symmetric component,
    distributed around the GC as the HG10 excess.}
  \label{fig:dm}
\end{figure}

\begin{figure*}
\includegraphics[width=.45\textwidth, angle=-90]{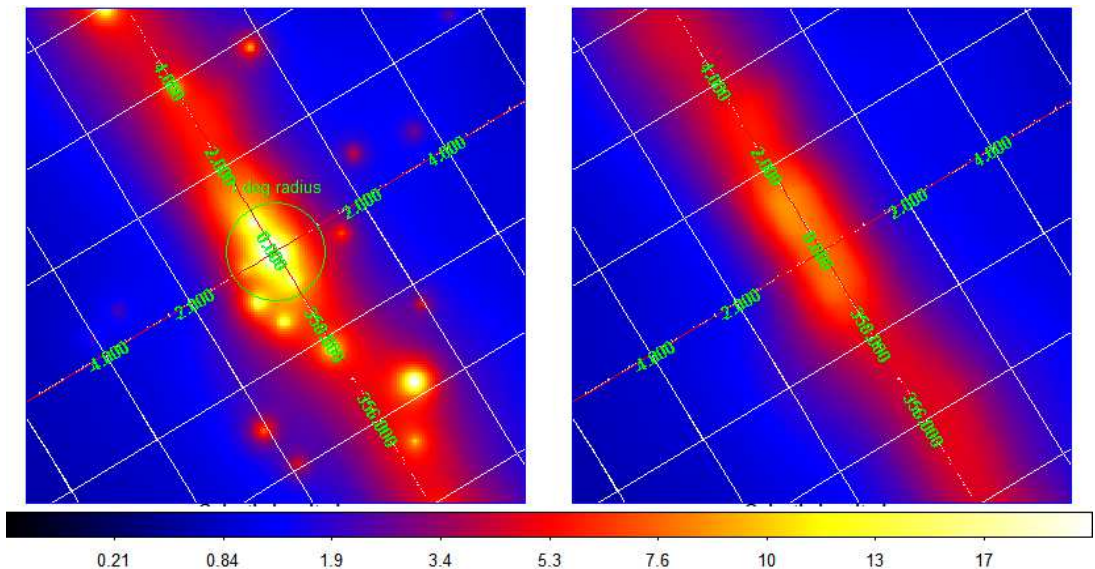}
\caption{Left: 10$^\circ$x10$^\circ$ count map of best-fit model. Right: only
  contribution from galactic and extragalactic backgrounds is shown}
\label{fig:model_cnt_map}
\end{figure*}

A different question however is whether such an additional component may be
ruled out. To this end we have added to our model of Sec.\ref{sec:Model} an
additional spherically symmetric component, whose intensity is distributed
around the center as $\rho^2(r)$ (where $\rho(r) \propto r^{-1.34}$, as found
in HG10).  We observe, that such a procedure does improve the fit (change in
the log-likelihood is 25 with only one new parameter added).  The resulting
spectral component is shown in Fig.~\ref{fig:dm}. Some of the photons from the
galactic diffuse background were attributed by the fit procedure to the new
component, concentrated in several central degrees (within the Galactic
Plane).  This phenomenon is probably related to the complicated and highly
non-uniform in the central region galactic diffuse background\footnote{See
  ``\textit{Description and Caveats for the LAT Team Model of Diffuse
    Gamma-Ray Emission}'' by the Diffuse and Molecular Clouds Science Working
  Group, Fermi LAT Collaboration,
  \url{http://fermi.gsfc.nasa.gov/ssc/data/access/lat/ring_for_FSSC_final4.pdf}.}
(cf. also the right panel of the Fig.~\ref{fig:model_cnt_map}).

We should also note that HG10 modeled diffuse background differently. They
considered contributions from the Galactic disk and spherically symmetric
emission in the region \emph{outside} central $2^\circ$ and then extrapolated
the diffuse model into the innermost $1^\circ-2^\circ$, arguing that the
contribution does not vary significantly in the range $2^\circ - 10^\circ$
off-center.  The background model we used (see \citealt{1FGLcat,FermiEGB} for
the detailed description) is different from that of HG10, especially in the central
1-2$^\circ$, where the model flux is higher than  the one extrapolated from larger
galactic longitudes, as one can clearly see on the right panel of the
Fig.~\ref{fig:model_cnt_map}.

\bigskip

Having the above considerations in mind, we think that the spectrum of the
central region, changing monotonously with the energy, is well described by
purely astrophysical model of the central point source and therefore present
data do not require any additional physical ingredients, such as DM
annihilation signal or %even
additional contributions from millisecond pulsars.  However, to firmly rule
out the emission from DM annihilation in the GC, more detailed model of the
galactic diffuse background is required.  Additionally, with the future data,
better statistics will reduce the error bars on the data point around $\sim
100$~GeV which will be helpful to better understand the central point source
physics.

\subsubsection*{Acknowledgments} We would like to thank M.~Chernyakova,
J.~Cohen-Tanugi, D.~Hooper, I.~Moscalenko, A.~Neronov, I.~Vovk for useful
comments. This work of A.B. and O.R. was supported in part by Swiss National
Science Foundation and by the SCOPES grant No.~IZ73Z0\_128040. The work of
D.M. was supported by grant 07/RFP/PHYF761 from Science Foundation Ireland
(SFI) under its Research Frontiers Programme.

\def\apj{ApJ}%
          % Astrophysical Journal
\def\apjl{ApJ}%
          % Astrophysical Journal, Letters
\def\apjs{ApJS}%
          % Astrophysical Journal, Supplement
\def\araa{ARA\&A}%
          % Annual Review of Astron and Astrophys
\def\aap{A\&A}%
          % Astronomy and Astrophysics

%\bibliography{hoop}
\end{document}